\documentclass[11pt]{article}
\usepackage{geometry}                
\usepackage{graphicx}
\usepackage{amssymb}
\usepackage{amsmath,amssymb}
\usepackage{epsfig}
\usepackage{graphicx}

\usepackage[usenames,dvipsnames]{color}

\newcommand{\be}{\begin{equation}}
\newcommand{\ee}{\end{equation}}
\newcommand{\bea}{\begin{eqnarray}}
\newcommand{\eea}{\end{eqnarray}}
\title{
\vspace*{-2cm}
\begin{flushright}
\normalsize{EFI-11-8 \\
ANL-HEP-PR-11-17\\ }
~\\
\end{flushright}
\vspace*{1.5cm}
Search for Higgs Bosons in SUSY Cascade Decays and Neutralino Dark Matter \\
\author{\textbf{Stefania Gori$^a$, Pedro  Schwaller$^{c,d}$ and Carlos E.M. Wagner$^{a,b,d}$} \\
~\\
\normalsize\emph{$^a$Enrico Fermi Institute \& $^b$Kavli Institute for Cosmological Physics,}\\
\normalsize\emph{University of Chicago, Chicago, IL 60637} \\
\normalsize\emph{$^c$ Physics Department, University of Illinois at Chicago, Chicago, IL 60607}\\
\normalsize\emph{$^d$HEP Division, Argonne National Laboratory, 9700 Cass Ave., Argonne, IL 60439}
}}
\begin{document}
\maketitle
\vspace*{0.5cm}
\begin{abstract}
The Minimal Supersymmetric Extension of the Standard Model (MSSM) is a well motivated theoretical framework, which contains an 
extended Higgs sector, including a light Higgs with Standard Model-like properties in most of the parameter space. Due to the 
large QCD background, searches 
for such a Higgs, decaying into a pair of bottom quarks, is very challenging at the LHC.  It has been long realized that the situation may be ameliorated by searching for Higgs bosons in supersymmetric
decay chains. Moreover, 
it has been recently suggested that the $b\bar{b}$ decay channel may be observed in standard production channels
by selecting boosted Higgs bosons, which may be easily identified from the QCD background.   Such boosted
Higgs bosons are frequent in the MSSM, since they are produced from decays of heavy colored supersymmetric particles.  Previous works have
emphasized the possibility of observing boosted Higgs bosons in the light higgsino region. 
In this work, we study the same question in the regions
of parameter space consistent with a neutralino dark matter relic density,
 analyzing its dependence on the non-standard Higgs boson, slepton
and squark masses, as well as on the condition of gaugino mass unification. In general, we conclude that, provided sleptons are heavier
than the second lightest neutralinos, the presence of boosted Higgs is a common  MSSM feature, implying excellent prospects for observation 
of the light MSSM Higgs boson in the near future.
\end{abstract}
\thispagestyle{empty}
\newpage
\section{Introduction}
The MSSM is a well motivated and extensively studied extension of the Standard Model (SM)~\cite{Haber:1984rc}--\cite{Martin:1997ns}.
 Its particle content is dictated by symmetry
and the couplings of all new particles are governed by the gauge and Yukawa couplings of the SM.  Among the most attractive properties
of the MSSM we can mention that it provides a renormalizable and perturbative theory, valid up to scales of the order of the Planck scale, it is consistent
with the unification of gauge couplings at high energies, it leads to a relation between the weak scale and the supersymmetric particle
masses and it contains a natural dark matter candidate, once  R-parity is implemented.  Moreover,  the MSSM contains a light Higgs, with a mass smaller than about 130~GeV~\cite{Haber:1990aw}--\cite{Martin:2003}, and SM-like properties in most of the parameter space. Searches for such a light Higgs boson
are of central importance since that particle is strongly linked to the mechanism of electroweak symmetry breaking. 

Searches for SM-like Higgs bosons have been performed at the LEP electron-positron collider at CERN as well as at the Tevatron collider
at Fermilab. LEP has established a lower bound on its mass of about 114.4~GeV~\cite{Barate:2003sz}--\cite{Schael:2006cr}, while the Tevatron has excluded the presence
of SM-like Higgs bosons with masses close to twice the $W$ mass~\cite{Abazov:2008eb}--\cite{:2009je}.  
In the low mass region, the Tevatron becomes most sensitive for masses
close to the LEP bound. At the end of its run, it is expected to reach a 3-$\sigma$ sensitivity for such low masses. 
After a full analysis and combination of the CDF and D0 data it will also probe most of the MSSM parameter space at the 2-$\sigma$ level~\cite{Draper:2009au}--\cite{Carena:2010ev}. The Tevatron has no discovery potential for a SM-like Higgs boson in this 
mass region. 

Discovery of such a light-Higgs boson, if it exists, is reserved to the LHC. Due to its SM-like properties, searches for the light MSSM Higgs boson 
at the LHC may proceed in the standard production channels, including
gluon-gluon fusion, with Higgs decaying into a pair of photons, as well as into neutral and charged gauge bosons, and weak boson fusion
with the Higgs decaying into a pair of tau leptons~\cite{Aad:2009wy}--\cite{Ball:2007zza}.  Associated production of the Higgs with top-quarks and $W^{\pm}$-bosons may
also be used at the LHC  after selecting the subset of boosted Higgs bosons~\cite{Butterworth:2008iy},\cite{Plehn:2009rk}. 

Preliminary analysis suggest that probing a very light Higgs, with a mass close to the LEP bound becomes challenging at the early
run of the  LHC, with a center of mass of 7~TeV, and will demand a few fb$^{-1}$ per experiment~\cite{Chamonix}, something
that is expected only by the shutdown at the end of 2012. Higgs discovery will be challenging in this region. The main search channel for
a SM-like Higgs at these energies and luminosities is the Higgs decay into two photons. Such a decay channel presents further
challenges in the MSSM, since its branching ratio tends to be suppressed due to a (small) mixing component of the 
light Higgs into non-standard Higgs bosons with enhanced couplings to bottom quarks and tau leptons. For these reasons, it
is very important to study alternative production channels.

In the MSSM, Higgs boson production may proceed from the decay of heavier supersymmetric particles.  Higgs produced
in the decay of squarks and gluinos are associated with hard jets (and leptons) and large missing energy, that allows
an effective suppression of the large QCD background.  Moreover, the Higgs bosons tend to be generally boosted. It has been 
recently suggested that boosted Higgs bosons may be easily identified from the QCD background even in standard production
channels~\cite{Butterworth:2008iy}. If they proceed from the decay of supersymmetric particles, such techniques can further enhance the probability of
observing a light Higgs boson~\cite{Kribs:2009yh}. 
Quite recently, a dedicated analysis of the possibility of observing boosted Higgs bosons in supersymmetric particle decays
was presented~\cite{Kribs:2010hp}. The authors concentrated mostly in the region of light higgsinos, where boosted Higgs
bosons are prominent. Such light higgsinos tend to lead to a low dark matter relic density due to the large higgsino
annihilation cross section. It is therefore interesting to study the possibility of observing (generally boosted) Higgs bosons in the regions
of parameters leading to  a neutralino density consistent with  the observed relic density.  

In this article, we perform such
a study, and find that in general the light Higgs boson can be observed in decay chains of supersymmetric particles, also in those regions of parameter space that are preferred by the requirement of obtaining the proper neutralino dark matter density. In section \ref{sec:two} we review the relic density constraints on the MSSM parameter space, and correlate it with the requirements of a large yield of Higgs bosons in sparticle decay chains, for universal gaugino masses, while in Sec.~\ref{sec:variations} we study the effects of more general MSSM particle spectra. 
In Sec.~\ref{sec:lhc} we simulate the LHC signals for some benchmark points, and comment on the prospects for discovery in the current 7~TeV run as well as for the future 14~TeV run of the LHC, before we present our conclusions in Sec.~\ref{sec:conclusions}.

\section{The MSSM with Heavy Sfermions}\label{sec:two}
We shall first consider a region of the MSSM parameter space in which both squarks and sleptons are heavy, with $m_{\tilde{q}} = m_{\tilde{\ell}} \equiv m_{\tilde{f}} \approx 1$~TeV\footnote{In this work we do not consider any splitting between the soft masses of the three generation squarks and sleptons. With $m_{\tilde{q}}$ and $m_{\tilde{\ell}}$ we indicate the common SUSY breaking squark and slepton soft masses, respectively. The physical masses of the three squarks and sleptons will then experience small splittings because of radiative corrections and third generation quark mass dependence of the squark mass matrices.}. Assuming gaugino mass unification at the GUT scale and a trivial flavor structure in the squark sector, the phenomenology of the model only depends  on five input parameters at the electroweak scale:
\begin{align}
	M_1\,,\quad \mu\,, \quad \tan\beta\,,\quad M_A\,\quad{\rm and} \quad m_{\tilde{f}}\,.
\end{align}
Besides the sfermion mass scale $m_{\tilde{f}}$, these are the bino mass $M_1$, the $\mu$ parameter, the ratio of the Higgs vacuum expectation values $\tan\beta$, and the mass of the pseudoscalar Higgs boson, $M_A$. 
The remaining gaugino masses are determined by the universality relation, which at the TeV scale roughly is given by
\begin{align}
	M_3 \simeq 3 M_2 \simeq 6 M_1\,.
\end{align}
Gaugino mass universality is a consequence of supersymmetric grand unification. In this paper we consider the MSSM as an effective theory, such that the above relation can in principle be broken. This possibility will be analyzed in Secs.~\ref{sec:nonuniversal},~\ref{sec:nonuniversal2}. 

Large squark masses are motivated by the current null results of direct searches for supersymmetric particles by the Tevatron and LHC experiments~\cite{Khachatryan:2011tk,Collaboration:2011qk}. This, together with constraints from electroweak precision tests and from the measurements of several flavor observables, suggests a relatively heavy colored SUSY spectrum. 

Slepton masses on the other hand are less constrained by experiments, and tend to be smaller than squark masses in explicit SUSY breaking scenarios, e.g. in mSUGRA. Our analysis does not depend significantly on the slepton mass scale, as long as they are heavier than $\tilde{N}_{2,3}$, i.e. above $\sim 500$~GeV. As emphasized before, in this section we shall consider the case in which sleptons are heavy,  $m_{\tilde\ell}=1$~TeV. The case of $m_{\tilde\ell}<500$~GeV will be discussed separately in Sec.~\ref{sec:sleptons}. 
\subsection{Neutralino Dark Matter}\label{sec:DM}
R-parity conservation requires superpartners to be created or destroyed in pairs, leading the lightest supersymmetric particle (LSP) to
be stable and hence a possible candidate for dark matter. In particular the lightest neutralino $\tilde N_1$ is often assumed to be the LSP, unless there is a lighter gravitino. In spite of the fact that the MSSM with R-parity provides naturally a good candidate of dark matter, the predicted relic abundance does not easily agree with the value obtained by WMAP, $\Omega h^2=0.1123\pm 0.0035$~\cite{Jarosik:2010iu}.

Assuming gaugino mass universality, the lightest neutralino is a linear combination of higgsino and bino states. As a consequence, the mass and composition of the LSP  depend mainly on $M_1$ and $\mu$. 

The correct relic density is obtained either by considering a heavily mixed LSP, or by providing resonant annihilation through the pseudoscalar Higgs $A^0$. In the other cases, annihilation is either too strong, when the LSP is mostly higgsino, or too weak, when it is mostly bino, to reproduce the relic density measured by WMAP. Note that, due to the large squark and slepton masses we assume here, there is no coannihilation region. 

\begin{figure}[p]
\center
\hspace{2cm}$M_A = 300$~GeV $\tan\beta = 10$ \hspace{2.6cm} $M_A = 300$~GeV $\tan\beta=50$ \\
\includegraphics[width=.45\textwidth]{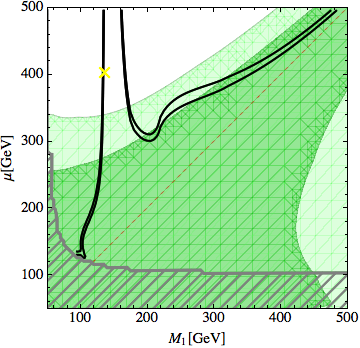}\hspace{0.6cm}\includegraphics[width=.45\textwidth]{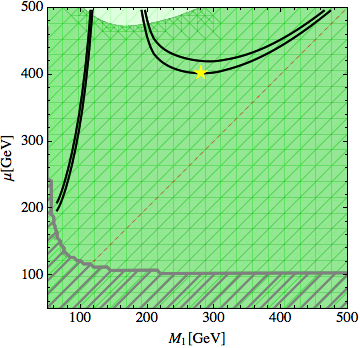}\\[.5em]
\hspace{1.8cm}$M_A = 1000$~GeV $\tan\beta = 10$ \hspace{2.5cm} $M_A = 1000$~GeV $\tan\beta=50$ \\
\includegraphics[width=.45\textwidth]{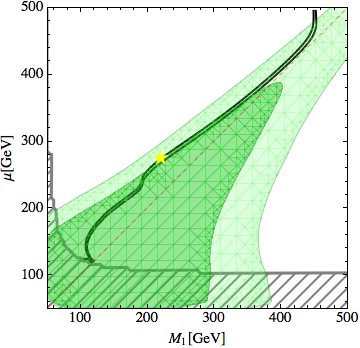}\hspace{0.6cm}\includegraphics[width=.45\textwidth]{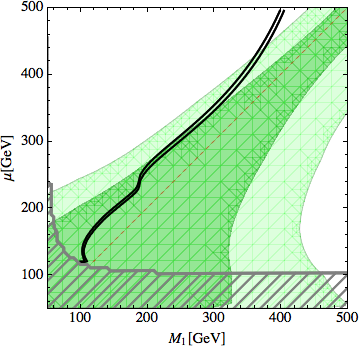}\\
\caption{\small Dark matter relic density in the  $M_1-\mu$ plane for heavy squarks and sleptons and $M_A=300~\rm{GeV}$ (top) and $M_A=1000~\rm{TeV}$ (bottom), for $\tan\beta = 10$ (left) and $\tan\beta = 50$ (right). 
The thin region between the solid black lines is the region in which the predicted relic density is in accordance with the experiments~\cite{Jarosik:2010iu}. The gray hatched region is excluded by LEP bounds on chargino masses. The green shaded regions are excluded by the latest Xenon~100 bounds on the spin independent dark matter-nucleon cross section, when using the most recent determination of the strange quark form factor $f_s=0.020$ (dark green) or the most conservative value for the strange quark form factor $f_s=0.118$ (light green). 
They yellow symbols denote benchmark points chosen for the collider analysis (see discussion in section~4).
}\label{Fig:Relic}
\end{figure}

In Fig.~\ref{Fig:Relic} we show the calculated relic abundance as a function of $M_1$ and $\mu$ for the two reference masses  $M_A=300~\rm{GeV}$ (top) and $M_A=1~\rm{TeV}$ (bottom) and for two different values of $\tan\beta=10,50$. The computation was performed using micrOMEGAs v2.4~\cite{Belanger:2010gh}. It can be noted that, independently on the value of $\tan\beta$, in the decoupling limit ($M_A\gg m_Z$) the preferred region is close to the line $M_1=\mu$, in which the LSP is a strongly mixed bino-higgsino state. Related studies were performed in~ \cite{ArkaniHamed:2006mb},\cite{Cirigliano:2006dg,Cirigliano:2009yd} (see also Ref.~\cite{Balazs:2004ae} for the case of light stops). 

For smaller $M_A$, the lightest neutralino can annihilate resonantly into heavy Higgs bosons, when $M_1 \sim M_A/2$. This provides a sufficiently large annihilation cross section for a mostly bino like LSP, such that the correct relic density can be obtained for $|\mu|\gg M_1$. The width of the resonant region increases for larger values of $\tan \beta$, due to the $\tan \beta$ enhancement of the coupling of $A$ to bottom quarks. Away from the resonant region, off-shell Higgs exchange still leads to an enhancement of the annihilation rate, thus favoring somewhat larger values of $\mu$ compared to the large $M_A$ case. Note that for $M_A=1~\rm{TeV}$ the tail of the resonance at $M_1 \sim 500~\rm{GeV}$ is already visible in Fig.~\ref{Fig:Relic}. Resonant annihilation can also be mediated by light Higgs boson exchange. The corresponding funnel region at $M_1 \sim m_h/2$ lies very close the parameter region that is excluded by direct LEP searches (see the gray hatched region in the figure), and is therefore not shown in the figures.

\bigskip

Dark matter direct detection experiments impose severe restrictions on the allowed parameter space. 
 For LSP masses $m_{\tilde{N}_1}>60$~GeV the most stringent constraints come from the CDMS~II~\cite{Ahmed:2009zw} and Xenon~100~\cite{Aprile:2011hi} experiments. 
Spin independent neutralino nucleon scattering is mediated by CP-even Higgs boson exchange. For large $\tan\beta$ and small $M_A$ the dominant contribution behaves as 
\begin{align}
	\sigma^{\rm SI} \sim \frac{\tan^2\beta}{M_A^4}\,,
\end{align}
where the $1/M_A^4$ dependence appears due to t-channel exchange of $H^0$ and the $\tan^2\beta$ behavior comes from the $\tan\beta$ enhanced couplings to down-type quarks. 
In particular, the potentially interesting region of small $M_A$ and large $\tan\beta$ is highly constrained. The theoretical uncertainty on the prediction for $\sigma^{\rm SI}$ is dominated by the strange quark form factor of the nucleon. Recent lattice studies~\cite{Ohki:2008ff,Ohki:2009mt} point to very low values of $f_s = 0.020 $ with $f_s< 0.08$ at the $1\sigma$ level, significantly smaller than the classical value~\cite{Ellis:2000ds} $f_s = 0.118\pm 0.062$ used in many previous analyses. For related discussions, see also~\cite{Giedt:2009mr,Cao:2010ph,Gogoladze:2010ch}. 

The excluded parameter region is shown green shaded in Fig.~\ref{Fig:Relic}. For WIMP masses of  $(50-300)$~GeV,
the latest results from Xenon~100~\cite{Aprile:2011hi} are up to a factor of four stronger than the previous combination of CDMS II and Xenon~100 limits, and now exclude a significant region of parameter space both for the small and large $M_A$ scenarios. In particular the so called well tempered neutralino region is only marginally compatible with direct detection constraints. 
To illustrate the effect of choosing different values for $f_s$, we show in the same plot the exclusion we would get for $f_s=0.020$ (dark green) and the one for $f=0.18$ (light green).

For smaller values of $M_A$ the constraints from direct detection start to exclude most of the relevant parameter space, in particular for large $\tan\beta$. However one should keep in mind that, in addition to the uncertainty coming from the strange quark form factor, dark matter direct detection constraints are also subject to astrophysical uncertainties like the local dark matter density and velocity distribution.  

\bigskip
In addition to direct detection experiments, neutralino dark matter is also constrained by experiments that are sensitive to products of neutralino annihilation in the sun or in the center of the galaxy.
The strongest constraints come from the SuperKamiokande and IceCube experiments that puts limits on high energy neutrinos produced in the sun~\cite{Abbasi:2009uz}. 
The most recent results from the Xenon~100 experiment however provides the strongest bounds on the region of parameter space relevant for our study. 

\subsection{Higgs Bosons From Neutralino and Chargino Decays}

The main source of Higgs bosons in supersymmetric decay chains are the two-body decays of neutralinos and charginos, 

\begin{align}
\tilde N_i&\rightarrow H_k \tilde N_j\,,\label{eqn:ndec}\\
\tilde C_2&\rightarrow H_k \tilde C_1\,,\notag
\end{align}
\noindent where $H_k$ is one of the three neutral Higgs bosons of the MSSM. In particular, since the lightest Higgs boson $h$ must have a mass below $130$~GeV, it is the most likely of the Higgs scalars to appear in these decays. 

The origin of these decay modes is the gauged kinetic term of the Higgs supermultiplets,
\begin{align}
	{\cal L} & = -D_\mu H_u^\dagger D^\mu H_u - i \bar{\tilde{H}}_u {D\!\!\!\!/} \tilde{H}_u - \sqrt{2} g' Y_{H_u} \tilde{B} \tilde{H}_u H_u^* - \sqrt{2} g \tilde{W}^a \tilde{H}_u t^a H_u^* + (u \leftrightarrow d)\,.
\end{align}
The neutralinos and charginos of the MSSM are linear combinations of the gauginos and higgsinos, 
\begin{align}
	\tilde{N}_i & = N_{i1} \tilde{B} + N_{i2} \tilde{W} + N_{i3} \tilde{H}_u + N_{i4} \tilde{H}_d \,,\\
	\tilde{C}_i & = C_{i1} \tilde{W}^+ + C_{i2}\tilde{H}^+\,.
\end{align}
The amount of mixing then determines which of the decay modes in (\ref{eqn:ndec}) have a large branching fraction, provided that they are allowed kinematically. The diagonalization of the neutralino and chargino mass matrices is straightforward, however the dependence on the MSSM parameters is nontrivial, thus it is more convenient to obtain the mixing matrices and branching fractions numerically. 

In Fig.~\ref{Fig:neutralinoandchargino} we show the branching fractions for the heavy neutralinos and the heavy chargino, in the $M_1-\mu$ plane, for $M_A=300$~GeV and $\tan\beta = 10$, considering only direct decays into Higgs bosons. The dependence on $\tan \beta$ and $M_A$ is rather weak, and mostly affects the region below the $M_1 = \mu$ line that is disfavored by the relic density. 
These results were obtained using the SUSY-HIT package~\cite{Djouadi:2007kx}. 

Most qualitative features of the branching fractions in Fig.~\ref{Fig:neutralinoandchargino} can be understood by looking at the composition and the mass spectrum of the relevant particles. 
The decay $\tilde{N}_2 \to \tilde{N}_1 h$ is only possible if both $M_2 - M_1 \simeq M_1 > m_h$ and $\mu -M_1 > m_h$ are satisfied, giving rise to the triangular shape in Fig.~\ref{Fig:neutralinoandchargino}(a). Since $\tilde{C}_1$ is close in mass to $\tilde{N}_2$ in this region, the only other allowed decay mode is $\tilde{N}_2 \to \tilde{N}_1 Z$. 
{In general, in this region of parameters, the Higgsino components of $\tilde N_1$, which is predominantly
a bino state,  are small and of opposite sign.  The Higgsino components of $\tilde N_2$ carry also opposite
signs and are small for $M_2 < \mu$. For $\mu < M_2$ instead, $\tilde N_2$ becomes approximately an
antisymmetric combination of the two Higgsinos.   Since the left- and right-handed coupling of neutralinos
to the Z  boson depend on the difference of the product of the up and down Higgsino components of both
neutralinos,
\begin{equation}
{N_{i3} N_{j3}^* - N_{i4} N_{j4}^*}\,,
\end{equation}
{while the couplings to the Higgs depend on the product of the gaugino and
Higgsino components, this  results in a suppression of the $\tilde N_2 \tilde N_1 Z$  coupling compared to the $\tilde N_2 \tilde N_1 h$
coupling. Therefore the decay $\tilde N_2\to \tilde N_1 Z$ is suppressed, and the Higgs branching fraction from $\tilde N_2$ decays
can reach 90\%.  }

{Differently, the orthogonal linear combination $\tilde N_3$ is  approximately a symmetric combination of Higgsinos
and hence in general the $\tilde N_3 \tilde N_1 Z$ coupling is large compared to the $\tilde N_2 \tilde N_1 Z$ one.  At the same time this
state acquires an axial coupling to $\tilde N_1 h$ (compared to the scalar $\tilde N_2 \tilde N_1 h$ coupling). Together this leads to a velocity
suppression  of the $\tilde N_3 \to h \tilde N_1$ decay and to a strong suppression of the branching ratio of $\tilde N_3$ decaying into Higgs bosons. }

The situation is inverted below the $M_1 = \mu$ line, as can be seen from Figs.~\ref{Fig:neutralinoandchargino}(a) and \ref{Fig:neutralinoandchargino}(b). Here $\tilde{N}_1$ and $\tilde{N}_2$ are higgsino-like, while $\tilde{N}_3$ is mostly bino and decays into Higgs-higgsino pairs. 
Furthermore also the decay $\tilde{N}_3 \to \tilde{C}_1^\pm W^\mp$ is possible, such that the Higgs boson branching fraction reaches at most 25\%, in accordance with the Goldstone boson equivalence theorem. On the other hand $\tilde{N}_1$ and $\tilde{N}_2$ have similar masses, such that decays into Higgs bosons are forbidden. 

In the region where $\tilde{N}_4$ is mostly wino, i.e. $M_2 =2 M_1 > \mu$, it decays into Higgs-higgsino pairs, which again gives a branching fraction to the light Higgs of at most 25\%. When the heavy Higgs states $H$ and $A$ become kinematically accessible, the branching fraction of $\tilde{N}_4$ into the light Higgs boson
is reduced accordingly. 
In the vicinity of $\mu = 2 M_1$ the decays $\tilde{N}_4 \to h \tilde{N}_{2,3}$ are kinematically forbidden, while $\tilde{N}_4 \to h \tilde{N}_1$ is suppressed by mixing. 
This can also be seen easily from Fig.~\ref{Fig:neutralinoandchargino}(c).

Finally the branching fraction of the chargino into the lightest Higgs boson follows a pattern similar to the one of $\tilde{N}_4$, as long as only direct decays are considered. This is shown in Fig.~\ref{Fig:neutralinoandchargino}(d). In addition, one should keep in mind that secondary Higgs bosons are produced from decays $\tilde{C}_2 \to \tilde{N}_{2,3}X$, in particular when $\tilde{N}_2 \to \tilde{N}_1 h$ is allowed. These secondary effects will be included in the analysis of squark decays. 

The little cascade  $\tilde{N}_2 \to \tilde{N}_1 h$ is a dominant source of Higgs bosons in the region $\mu > M_1 + m_h$. This is quite different from the scenarios that were considered in~\cite{Kribs:2010hp}, where this decay is mostly irrelevant. For smaller slepton masses, $m_{\tilde\ell}< m_{\tilde{N}_2}$, this channel will be depleted, as discussed in more detail in Sec.~\ref{sec:sleptons}. 

The heavier Higgs bosons $H^0$ and $A^0$ can in principle also appear in these decays. For the parameter ranges that are considered here, these decays are often phase space suppressed, if not forbidden. Exceptions are discussed in Secs.~\ref{sec:heavyuni} and \ref{sec:nonuniversal2}. 

Besides neutralino/chargino decays, Higgs bosons can also appear in the decay $\tilde{t}_2 \to \tilde{t}_1 h$ due to the large top quark Yukawa. In practice it turns out that this decay has a branching fraction of at most a few percent in the parameter regions studied here. 

\begin{figure}
\center%
\hspace{.04\textwidth}(a)\hspace{.2\textwidth}$(\tilde N_2 \to h\tilde N_1)$\hspace{.1\textwidth}(b)\hspace{.2\textwidth}$(\tilde N_3 \to h\tilde N_i)$\\
\includegraphics[width=.46\textwidth]{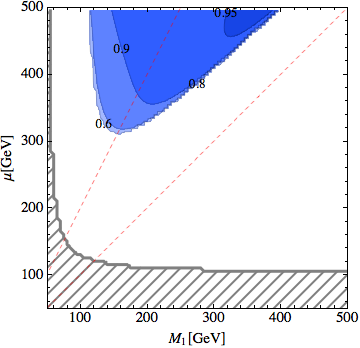}\hspace{.4cm}
\includegraphics[width=.46\textwidth]{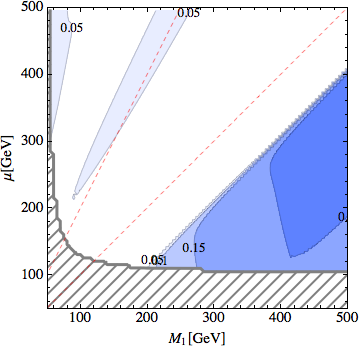}\\
\hspace{.04\textwidth}(c)\hspace{.2\textwidth}$(\tilde N_4 \to h\tilde N_i)$\hspace{.1\textwidth}(d)\hspace{.2\textwidth}$(\tilde C_2 \to h\tilde C_1)$\\\includegraphics[width=.46\textwidth]{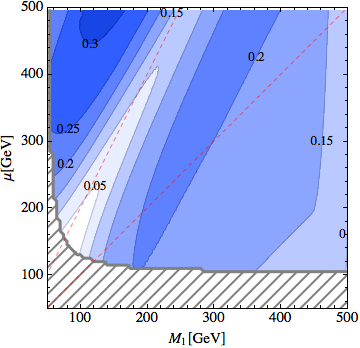}\hspace{.4cm}
\includegraphics[width=.46\textwidth]{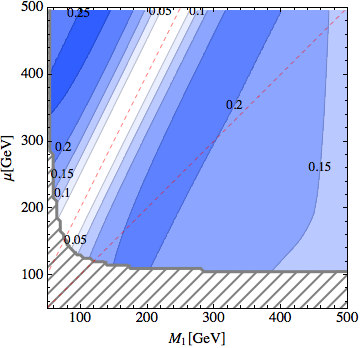}
\caption{\small 
Branching fractions for the direct decays of $\chi_i \to h+\chi_j$, where $\chi_i$ denotes both chargino and neutralinos, for $M_A = 300$~GeV and $\tan\beta = 10$. The branching fractions increase from light to dark blue, as indicated in the figures. The $\mu=M_1$ and $\mu = 2M_1 = M_2$ lines (dashed, red) are shown for easier orientation. The gray hatched parameter region is excluded by direct searches for charginos at LEP. }
\label{Fig:neutralinoandchargino}
\end{figure}
\subsection{Constraints on the Parameter Space}

Before analyzing the concrete possibility of producing the light Higgs boson through SUSY decay chains, in this section we study the constraints we have to impose to our $M_1-\mu$ plane. The main constraints on the parameter space originate from chargino searches at LEP, that impose lower bounds on the masses of $\tilde{N}_2$ and $\tilde{C}_1$ (see the gray hatched area in Fig.~\ref{Fig:neutralinoandchargino}). In addition, we require that the LSP is a neutral, color-singlet state, and that the contribution to the T parameter is sufficiently small to not upset electroweak precision constraints. To avoid negative contributions to $(g-2)_{\rm muon}$ we further restrict ourselves to positive values of $\mu$.

Finally, assuming a trivial flavor structure in the squark mass matrices, flavor constraints are rather mild in the region of parameter space we are analyzing. Potentially, the only relevant flavor observables that could get sizable new physics contributions are $b\rightarrow s\gamma$, $B\rightarrow\tau\nu$ and $B_s\rightarrow\mu^+\mu^-$. 

Concerning the branching ratio of $b\rightarrow s\gamma$, the dominant SUSY contributions arise from penguin diagrams with charged Higgs up-type quarks and chargino up-type squarks: for a pseudoscalar mass $M_A$ larger than $300$ GeV and squark masses of the order $1$ TeV both contributions are not too large and the resulting branching ratio is compatible with the experimental bound. Lowering the value of $M_A$, the allowed parameter space gets more constrained since the (always positive) Higgs contribution decouples as $1/M_A^2$. One needs then a sizable negative contribution coming from the chargino. Negative trilinear terms $A_t$ and products $A_t\tan\beta$ rather large are then favored, although the potential contribution from small flavor violating squark-quark-gluino couplings may also be important~\cite{Bertolini:1990if,Carena:2008ue}.  We will assume that constraints from flavor physics are satisfied. For definiteness we assume $A_t = - 1000$~GeV for the remainder of the paper. 

Constraints coming from $B\rightarrow\tau\nu$ and $B_s\rightarrow\mu^+\mu^-$ are also easily satisfied for a pseudoscalar mass of at least $300$ GeV.  Below that value the two branching ratios should be evaluated more carefully. Still, even for $M_A=200$ GeV, the experimental constraints can be satisfied choosing a not too large value of $\tan\beta$. The careful analysis of the flavor constraints is beyond the scope of this work, since it will not affect the main features discussed in this paper. 
\subsection{Higgs Production through Squark Decay Chains}
At the LHC, neutralinos and charginos can be directly produced. The cross sections can be sizable, up to $1~$pb, when they are sufficiently light. One possible signal, resulting from pair production of $\tilde{N}_2$, is a pair of Higgs bosons decaying to four b~quarks and missing energy. These signals are however very hard to disentangle from the large QCD background. 

It is more promising to look for Higgs bosons in decay chains of squarks and gluinos. Here the strong QCD production will lead to sizable rates even for large squark and gluino masses. Moreover, recently it has been pointed out that highly boosted Higgs bosons, originating from such decay chains, provide a handle to reduce the notorious QCD background for $h\to \bar{b}b$ decays, using jet-substructure techniques~\cite{Kribs:2010hp}. 

Since gluinos mainly decay to squark-quark pairs, the fraction of sparticle cascades that contain a Higgs boson is mostly determined by the probability for a squark decay to produce a Higgs boson. As a first approximation, this is given by the branching fraction of a squark into a given neutralino or chargino, multiplied by the probability that the neutralino or chargino decays into a Higgs boson:
\begin{align}
	{\rm P}(\tilde{q} \to h + X) & = \sum_{\chi_i} {\rm Br}(\tilde{q} \to \chi_i + q) \times {\rm Br}(\chi_i \to h + \chi_j)\,,
\end{align}
where $\chi_i$ denotes either a neutralino or a chargino. For the numerical results, we also include secondary effects, i.e. when the Higgs originates from a decay chain $\tilde N_3 \to \tilde N_2 X \to h \tilde N_1 X$, or similar.  

Gluinos with masses below $m_{\tilde{q}}$ decay into a neutralino or chargino and two quarks, mediated by an off-shell squark. Since we assume  $m_{\tilde{q}}=1$~TeV and gaugino mass universality here, this only affects the region where $M_1\lesssim 150$~GeV. The $\tilde{N}_2\to \tilde{N}_1 h$ decay is forbidden for $M_1\lesssim 120$~GeV, while the heavier neutralinos and charginos only appear very rarely. Therefore gluino three body decays yield Higgs bosons only in a very limited region of parameter space, while outside of this region the $\tilde{q}\to \tilde{g} q$ decays suppress the appearance of Higgs bosons. 

Before going to the numerical results for $P(\tilde{q} \to h + X)$, we can try to understand some general features that we expect to find.  
Left-handed squarks decay mostly into a quark and a wino, while the right-handed squarks decay almost exclusively into quarks and binos. The third generation squarks in addition can decay to quarks and higgsinos. From this, we can already deduce the main sources of Higgs bosons. To obtain the correct relic density, we are bound to a region where $|\mu| > M_1$, such that the LSP is mostly bino. This already excludes right-handed squarks as a relevant source of Higgs bosons. 
Away from the resonance region, the $\tilde{N}_2$ is mostly higgsino, such that the $\tilde{N}_2$ decays will greatly enhance the fraction of Higgs bosons in stop decays whenever they are kinematically accessible. In comparison, the parameters chosen in~\cite{Kribs:2010hp} for the boosted Higgs analysis focus on the $M_1 > \mu$ region, where Higgs bosons are produced both in the decays of winos and binos, when kinematically allowed, and thus
the contribution of right-handed squarks is comparable to the one of the left-handed squarks.

In Fig.~\ref{fig:combined2} we present our results for the branching ratios of the decay of first two generation up squarks (first row) and stops (second row) into the lightest Higgs boson, for  $M_A=1000~\rm{GeV}$. These branchings are largely independent of the value of $\tan\beta$. 
The corresponding branching fractions for down-type squarks are very similar to those of the first and second generation up-type squarks,  hence we do not present separate plots for their branching fractions. The colored bands indicate the region where the relic density agrees with observations for $\tan \beta = 10$ (black)  and $\tan\beta=50$ (green). Dark matter direct detection constraints are not imposed upon the parameter space, since they would be different for the different $\tan\beta$ contours shown. For the large $\tan\beta$ regime, where these constraints are most restrictive, allowed points can easily be identified by comparing with Fig.~\ref{Fig:Relic}. 

\begin{figure}
\center
\hspace{.04\textwidth}(a)\hspace{.2\textwidth}$(\tilde{q}_L \to h+X)$\hspace{.1\textwidth}(b)\hspace{.2\textwidth}$(\tilde{q}_R \to h+X)$\\
\includegraphics[width=.46\textwidth]{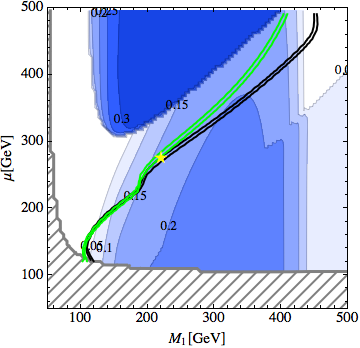}\hspace{0.4cm}\includegraphics[width=.46\textwidth]{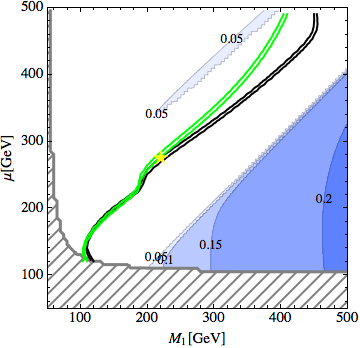}\\
\hspace{.04\textwidth}(c)\hspace{.2\textwidth}$(\tilde{t}_1 \to h+X)$\hspace{.1\textwidth}(d)\hspace{.2\textwidth}$(\tilde{t}_2 \to h+X)$\\
\includegraphics[width=.46\textwidth]{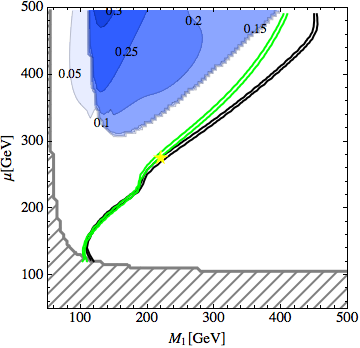}\hspace{0.4cm}\includegraphics[width=.46\textwidth]{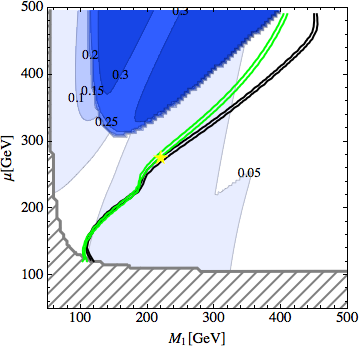}
\caption{\small Probability for a Higgs boson in squark decay chains, for $M_A=1000$~GeV. From lightest to darkest blue, the probabilities are 5\%, 10\%, 15\%, 20\%, 25\%, 30\%. The gray hatched area is excluded by LEP. Superimposed are the regions of correct relic density for $\tan\beta = 10$ (black) and $\tan \beta = 50$ (green). The constraints from dark matter direct detection are not shown. The yellow star indicates the benchmark point (I) discussed in Sec.~\ref{sec:lhc}. 
}\label{fig:combined2}
\end{figure}

It is clear from the plots that for large $M_A$ the only relevant source of Higgs bosons are the left-handed squark decay chains. Stop decays fail to generate a sizable contribution since the important $\tilde{N}_2 \to h \tilde{N}_1$ decay is inaccessible in the region where the relic density is correct. 

Lowering $M_A$ does not dramatically change the squark branching fractions, however the relevant region in parameter space gets shifted to higher values of $\mu$. We illustrate this for the particular case of $M_A = 300$~GeV in Fig.~\ref{fig:combined3}. 
In the vicinity of the resonance, both the left-handed squarks and the stops give a sizable amount of Higgs bosons. Away from the resonance, stop decays can only contribute for large values of $\tan \beta$ (see green band).

\begin{figure}
\center
\hspace{.04\textwidth}(a)\hspace{.2\textwidth}$(\tilde{q}_L \to h+X)$\hspace{.1\textwidth}(b)\hspace{.2\textwidth}$(\tilde{q}_R \to h+X)$\\
\includegraphics[width=.46\textwidth]{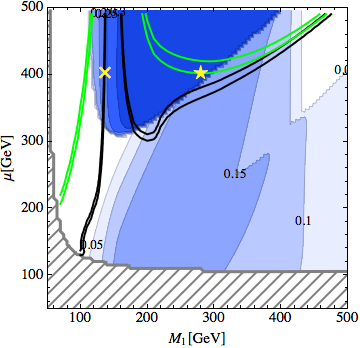}\hspace{0.4cm}\includegraphics[width=.46\textwidth]{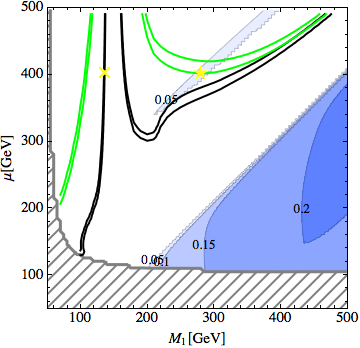}\\
\hspace{.04\textwidth}(c)\hspace{.2\textwidth}$(\tilde{t}_1 \to h+X)$\hspace{.1\textwidth}(d)\hspace{.2\textwidth}$(\tilde{t}_2 \to h+X)$\\
\includegraphics[width=.46\textwidth]{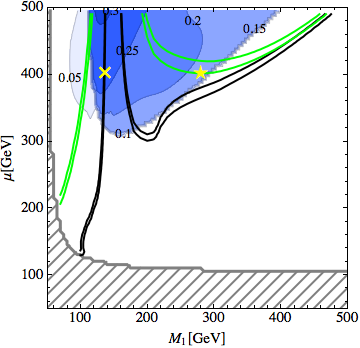}\hspace{0.4cm}\includegraphics[width=.46\textwidth]{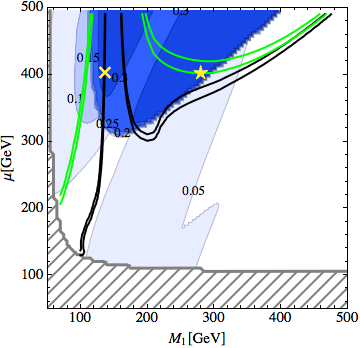}
\caption{\small Same as Fig.~\ref{fig:combined2}, for $M_A=300$~GeV. The yellow star and cross show the positions of the benchmark points (II) and (III) respectively, discussed in Sec.~\ref{sec:lhc}. 
}\label{fig:combined3}
\end{figure}

\subsection{Heavy Higgs Bosons from Squark Decays}\label{sec:heavyuni}

It is  important to concentrate on the region with small pseudoscalar masses $M_A$, since here it might be possible to also observe the heavier Higgs bosons, which tend to be difficult to observe for moderate or small values of $\tan\beta$, where their direct production cross section is small. 
In particular in the regions of parameter space where the neutralinos are sufficiently split in mass, squark decays can also lead to a sizable production of the heavy Higgs boson $H$ and of the pseudoscalar $A$. 

In Fig.~\ref{Fig:heavyuniversal} we present our results for the summed decay rates to the scalar and pseudoscalar heavy Higgs bosons, having fixed $M_A=200~{\rm GeV}$ and $\tan\beta=10$. We do not show the contribution of the right handed squarks of first two generations $\tilde q_R$ since they do not bring a significant branching ratio. The two stops give similar contributions, with slightly larger branching fractions for the heavier $\tilde{t}_2$.

\begin{figure}[t,h]
\center
\hspace{.04\textwidth}(a)\hspace{.16\textwidth}$(\tilde{q}_L \to H/A+X)$\hspace{.1\textwidth}(b)\hspace{.16\textwidth}$(\tilde{t}_2 \to H/A+X)$\\
\includegraphics[width=.46\textwidth]{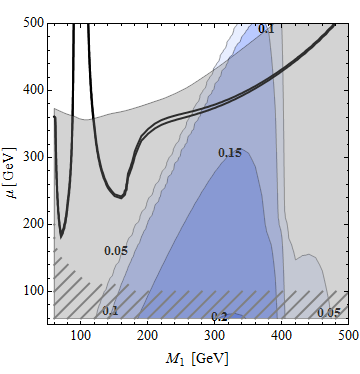}\hspace{0.4cm}\includegraphics[width=.46\textwidth]{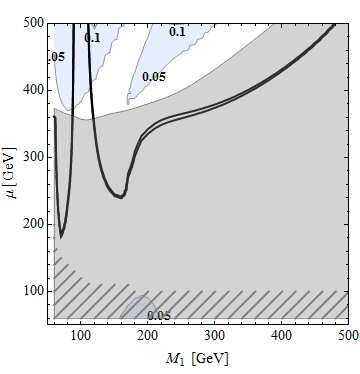}\\
\caption{\small Probability for a heavy Higgs boson or a pseudoscalar in squark decay chains, for $M_A=200$~GeV and $\tan\beta=10$. From lightest to darkest blue, the probabilities are 5\%, 10\%, 15\%. The gray hatched area is excluded by LEP. Superimposed are the regions of correct relic density and the region excluded by dark matter direct detection (shaded in gray). The most recent determination of the strange quark form factor $f_s=0.020$ has been used.}\label{Fig:heavyuniversal}
\end{figure}

As we can observe from the figure, for universal gaugino masses, heavy Higgs boson production is not favored. The stop branching ratios can reach $\sim 10\%$ in regions that are compatible with the requirement of a correct relic abundance and with dark matter direct searches. Differently, first two generation left-handed squarks do not contribute significantly to the heavy Higgs boson production in those regions.

 Dropping the universality relation between gaugino masses can lead to an enhanced production of heavy Higgs bosons in squark decay chains. This possibility is explored further in Sec.~\ref{sec:nonuniversal2}.

\section{More General MSSM Spectra}\label{sec:variations}
\subsection{Lighter Sleptons}\label{sec:sleptons}
The presence of lighter sleptons can reduce significantly the branching fractions for production of Higgs bosons in squark decay chains. In this section we will investigate the effects of lowering the slepton mass scale on the Higgs branching fractions. 

In Fig.~\ref{Fig:sleptons}, we restrict our attention to the case $M_A=300~\rm{GeV}$ and $\tan\beta=10$, and we compare the prediction for the dark matter relic abundance and the production of the lightest Higgs boson through the decay of a squark of the first two generation (left panels) and of a stop (right panels), arising in scenarios with different slepton masses. Squarks are always assumed to be heavy ($1~\rm{TeV}$). Lowering the slepton mass, a larger region of the $M_1-\mu$ plane is excluded because of the appearance of a stau LSP (gray hatched area for large values of $M_1$ and $\mu$). Note however that below $\mu\sim 450$~GeV the sneutrino becomes lighter than the stau, such that this parameter region is not excluded a priory. However the large sneutrino-stau co-annihilation rate strongly suppresses the sneutrino relic density, so that the sneutrino LSP region is not phenomenologically relevant.

\begin{figure}[p]
\center
\hspace{.04\textwidth}(a)\hspace{.2\textwidth}$(\tilde{q}_L \to h+X)$\hspace{.1\textwidth}(b)\hspace{.2\textwidth}$(\tilde{t}_2 \to h+X)$\\
\includegraphics[width=.46\textwidth]{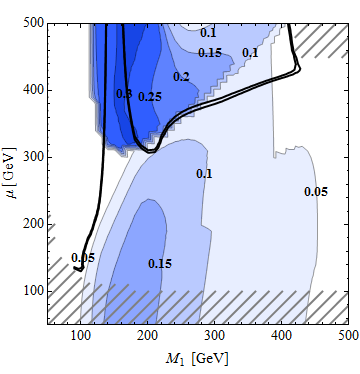}\hspace{0.4cm}\includegraphics[width=.46\textwidth]{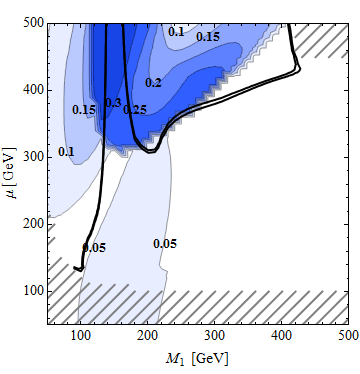}\\
\vspace{0.5cm}
\hspace{.04\textwidth}(c)\hspace{.2\textwidth}$(\tilde{q}_L \to h+X)$\hspace{.1\textwidth}(d)\hspace{.2\textwidth}$(\tilde{t}_2 \to h+X)$\\
\includegraphics[width=.46\textwidth]{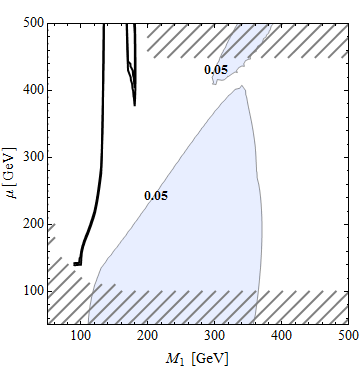}\hspace{0.4cm}\includegraphics[width=.46\textwidth]{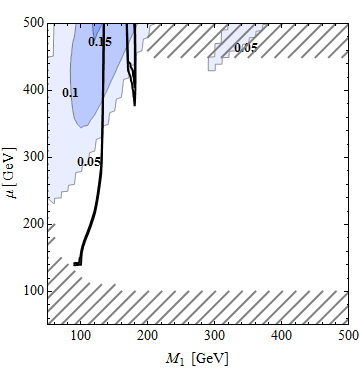}
\caption{\small Probability for a Higgs boson in squark decay chains, for
$M_A=300$ GeV and $\tan\beta=10$ and two different values for the common soft SUSY breaking slepton mass: $m_{\tilde\ell}=400~\rm{GeV}$ (first row) and $m_{\tilde\ell}=200~\rm{GeV}$ (second row). From lightest to darkest blue, the probabilities are $5\%$, $10\%$, $15\%$, $20\%$, $25\%$ and $30\%$. The gray hatched area is either excluded by LEP (at small values of $\mu$) or excluded by a stau LSP (at large values of $\mu$). Superimposed are the regions of correct relic density.
}\label{Fig:sleptons}
\end{figure}

From the figure, it is evident that the dark matter relic abundance only marginally depends on the slepton mass scale, as long as one stays away from the region where the stau becomes the LSP. Close to this region, the neutralino-stau co-annihilation visibly suppresses the relic density. 

A different behavior is shown by the branching ratios. When either the wino mass or the higgsino mass is larger than $m_{\tilde{l}}$, the decay channels into slepton lepton pairs open up, and thus reduce the Higgs branching fractions. 

For $m_{\tilde\ell}=400$ GeV, comparing Fig.~\ref{Fig:sleptons}(a,b) with the case of heavy sleptons (Fig.~\ref{fig:combined3}(a,d)), one can note that the Higgs probabilities are reduced only on right of the $M_1 = 200$~GeV line, where the wino mass is larger than the slepton mass. In this region the branching fraction to Higgs bosons is reduced roughly by a factor of two. In the region where in addition $\mu > 400$~GeV also the higgsino Higgs decays are depleted, such that the branching fraction to Higgs bosons is suppressed by a factor of three or more compared to the case of heavy sleptons. 

 For smaller slepton masses  ($m_{\tilde\ell}=200\,\rm{GeV}$) and large values of $\mu$, the contribution of squarks of the first two generation is tiny, since the next to lightest neutralino decays mainly into lepton-slepton pairs. Differently, the stops still contribute to the production of the lightest Higgs boson, since the heaviest neutralino $\tilde N_4$ has still sizable branching fractions into the lightest Higgs. Assuming stops have similar masses as the first and second generation squarks, due to the relatively small stop quark production cross section at LHC, it will be very difficult to observe a Higgs boson in this very light slepton scenario.
 
 While here we have assumed equal soft masses for the left- and right-handed sleptons, the main suppression of the Higgs production is due to winos decaying to left-handed sleptons. If only the masses of the right-handed sleptons are lowered, then the sleptonic decay modes of the wino are suppressed by the mixing of left- and right-handed sleptons, and dominated by decays into stau-tau. Due to the reduced number of accessible final states, in this case we expect that a sizable branching fraction of neutralinos into Higgs bosons survives.

 \bigskip
From the above discussion, we find the following condition for a sizable production of the light Higgs boson in squark decay chains: 

\begin{align}
	m_{\tilde\ell} > M_2 = 2 M_1 > 2 m_h\,.
\end{align}

The condition prevents in fact wino decays to sleptons. 
If  $\mu > m_{\tilde\ell}$, then the Higgs bosons will mostly be produced in the decay $\tilde{N}_2 \to \tilde{N}_1 h$. In order to satisfy the relic density constraint in this regime, we have to require the additional condition $M_A \approx 2 M_1$, which implies that also $m_{\tilde\ell}>M_A$. 
On the other hand, if both $\mu < m_{\tilde\ell}$ and $m_{\tilde\ell}> 2 M_1$, we essentially recover the case of heavy sleptons. 
\subsection{Nonuniversal Gaugino Masses}\label{sec:nonuniversal}
The universal relations between the gaugino masses at the electroweak scale, 
\begin{align}
	6 M_1 \approx 3 M_2 \approx M_3\,,
\end{align}
may be a consequence of grand unification, or of a supersymmetry breaking mechanism that depends on gauge interactions, e.g. in minimal gauge mediation. 
 From the point of view of a low energy effective theory, there is no relation between these parameters, and thus they should be treated independently. 
In addition, even in the context of grand unification, these relations are modified if non-singlet $SU(5)$ chiral superfields appear in the gauge kinetic function. Possible representations have been studied e.g. in~\cite{Huitu:2008sa}, and can lead to completely different relations between the gaugino masses at the electroweak scale.  The impact of general gaugino masses on Higgs production has also been considered in~\cite{Bandyopadhyay:2008fp,Bandyopadhyay:2008sd}, and a recent study in the context of neutralino dark matter can e.g. be found in~\cite{Vasquez:2010ru}.

We assume that $M_3$ is of the order of the squark mass parameter (1 TeV) to avoid constraints from direct searches (see e.g~\cite{Khachatryan:2011tk,Collaboration:2011qk} for recent updates). The regime where $M_2$ is very close or even smaller than $M_1$ is disfavored since the relic density would be strongly suppressed. 

On the other hand, increasing the ratio $M_2/M_1$ is possible without any evident difficulty. One immediate consequence is that the lower bound on $M_1$ coming from LEP is relaxed, and the mass of the lightest neutralino can be lowered.  In this very light neutralino regime, large enough annihilation cross sections are only obtained close to the $Z$ and $h$ resonances ($m_{\tilde{N}_1} \approx m_Z/2$ or $m_h/2$), requiring some fine tuning among the mass parameters.\footnote{Additional annihilation channels involving light sleptons are absent in our scenarios.} In these resonant regions, the probability to find Higgs bosons in $\tilde{q}_L$ and $\tilde{t}_{1,2}$ decays can reach $20\%$ or more. An example is shown in Fig.~\ref{fig:smallm1} for $M_2 = 400$~GeV, $M_A=300$ GeV and $\tan\beta=10$. Due to the small LSP mass, this region is particularly interesting for a boosted Higgs search. Direct detection experiments do not constrain the parameter space in the regime where the relic density is compatible with observations. 

\begin{figure}[t]
\center
\hspace{.04\textwidth}(a)\hspace{.2\textwidth}$(\tilde{q}_L \to h+X)$\hspace{.1\textwidth}(b)\hspace{.2\textwidth}$(\tilde{t}_1 \to h+X)$\\
\includegraphics[width=.46\textwidth]{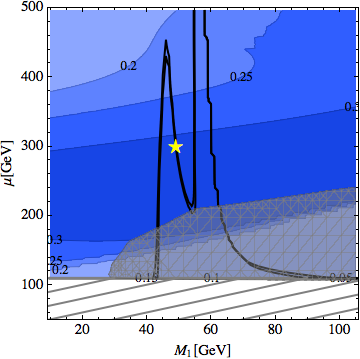}\hspace{.6cm}
\includegraphics[width=.46\textwidth]{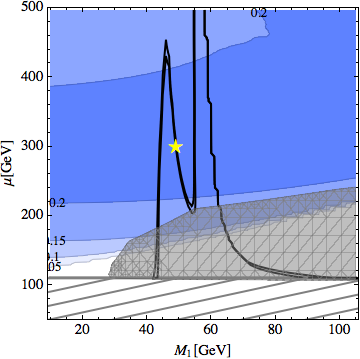}
\caption{Shown are the regions of correct relic density (black band) and the probability for finding a Higgs boson in $\tilde{q}_L$ (left) and $\tilde{t}_1$ (right) decay chains, for a scenario with nonuniversal gaugino masses, with $M_2  = 400$~GeV, $M_3 = 1$~TeV, $M_A = 300$~GeV and $\tan\beta=10$. As before, the gray hatched area is excluded by direct searches at LEP  and the gray shaded area is excluded by Xenon~100. The yellow star indicates the benchmark point (IV) discussed further in Sec.~\ref{sec:lhc}.
}
\label{fig:smallm1}
\end{figure}

\subsection{Heavy Higgs Boson Production for Nonuniversal Gaugino Masses}\label{sec:nonuniversal2}
Let us recall that in the case of universal gaugino masses, 
in order to obtain a large branching fraction into heavy Higgs bosons, the neutralino mass splittings must be larger than $M_A$. On the other hand the relic density constraint implies that either $\mu \sim M_1$ or $M_1 \sim M_A/2$. 
In the former case, a large enough mass splitting is obtained for $M_1 \gg M_A$, as we found in Sec.~\ref{sec:heavyuni}. In the latter case, the neutralinos can be made sufficiently heavy by increasing $\mu$, however in this case the heavy neutralinos are mostly higgsino like and do not appear in decays of the first and second generation squarks. 

Dropping the requirement of universal gauging masses, these constraints can easily be circumvented. In particular it is obvious that increasing the ratio $M_2/M_1$ will allow us to obtain heavy Higgs bosons from wino decays also for smaller values of $M_1$. 

As an example, in Fig.~\ref{Fig:heavynonuniversal} we present our investigation for the summed decay rates to the heavy Higgs boson and to the pseudoscalar for the particular case the gaugino mass ratios are determined by the {\bf{24}} representation of $SU(5)$

\begin{equation}
14M_1\approx 2.3 M_2\approx -M_3\,.
\end{equation}

Thanks to the increased mass ratio $M_2/M_1$, the second lightest neutralino can make a sizable contribution to the heavy Higgs production. The production rates are enhanced compared to the universal gaugino mass case (see Fig.~\ref{Fig:heavyuniversal} for comparison), as also noticed in~\cite{Huitu:2008sa}. For $M_1$ slightly larger than $100$ GeV and the $\mu$ parameter rather large (around $400$-$500$ GeV), the branching fractions can reach the $25\%$ level both from first generation left-handed squark and stop decays, in regions compatible with a correct DM relic abundance and with constraits from DM direct detection. This would certainly improve the chances of an observation of the heavy Higgs states at the LHC.

\begin{figure}[t]
\center
\hspace{.04\textwidth}(a)\hspace{.16\textwidth}$(\tilde{q}_L \to H/A+X)$\hspace{.1\textwidth}(b)\hspace{.16\textwidth}$(\tilde{t}_2 \to H/A+X)$\\
\includegraphics[width=.46\textwidth]{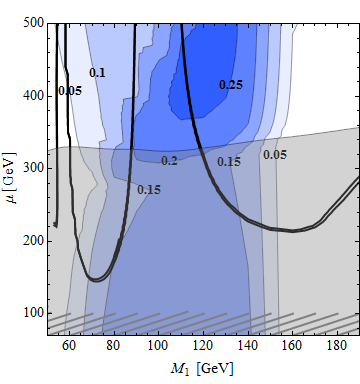}\hspace{0.4cm}\includegraphics[width=.46\textwidth]{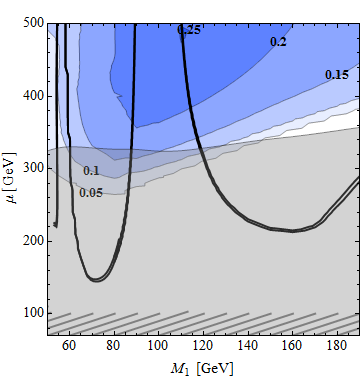}\\
\caption{\small Probability for a heavy Higgs boson or a pseudoscalar in squark decay chains, for $M_A=200$~GeV and $\tan\beta=10$, assuming gaugino mass ratios dictated by the {\bf{24}} representation of $SU(5)$. From lightest to darkest blue, the probabilities are 5\%, 10\%, 15\%, 20\%, 25\%. The gray hatched area is excluded by LEP. Superimposed are the regions of correct relic density and the region excluded by dark matter direct detection (shaded in gray).}\label{Fig:heavynonuniversal}
\end{figure}

The ratios of wino and bino masses obtained for a {\bf{24}} representation of $SU(5)$ seems to be a very good choice to obtain large heavy Higgs production rates. Larger ratios may be considered, but do not lead to a significant improvement over the  {\bf{24}} scenario, analyzed above.

\subsection{Charged Higgs Production}

Before closing this section, let us also mention that charged Higgs production can be enhanced in scenarios with nonuniversal gaugino masses. For the universal case, the branching fractions into $H^+$ are suppressed similarly to the neutral heavy Higgs bosons, and are below 10\% for most of the parameter space. In particular the decay $\tilde{C}_1 \to \tilde{N}_1 H^+$ is hardly possible. 

Increasing the mass splitting between the wino and the bino, $M_2 > M_1 + M_A$, and assuming $\mu \gtrsim M_2$, the decays $\tilde{C}_1 \to \tilde{N}_1 H^+$ become kinematically accessible. The only competing decay mode is $\tilde{C}_1 \to \tilde{N}_1 W^+$. If the decay to the charged Higgs is not phase space suppressed, it will have a branching fraction of up to 50\%.  

The lightest chargino is abundant in left-handed squark decays whenever $\tilde{C}_1$ has a large wino component. Assuming the above relations between the neutralino mass parameters, the probability to observe a charged Higgs boson in left-handed squark decays can reach up to 25\%. 

We will not attempt here to estimate the possibility of observing the decays of $H^+$ into top bottom pairs from these production channels. However, due to the small neutralino mass, most charged Higgs bosons will be boosted significantly, with transverse momenta of 300~GeV or more. It may therefore be worth to study the applicability of boosted Higgs searches also to charged Higgs boson decays to top-bottom pairs. 

\section{Higgs Cascades at LHC}\label{sec:lhc}
The possibility to look for the Higgs in MSSM decay chains has  been studied in the past, e.g. in~\cite{Baer:1992ef,Datta:2001qs,Datta:2003iz,Bandyopadhyay:2008fp,Huitu:2008sa,Bandyopadhyay:2010tv}. In these studies mostly conventional cut based analyses are used to distinguish the Higgs signal from background events and to estimate the statistical significance. More recently it has been suggested that techniques based on jet substructure algorithms can be used to improve the signal to background ratio, provided that at least a fraction of events contain a Higgs boson with a large transverse momentum~\cite{Kribs:2009yh,Kribs:2010hp}.
Both conventional and subjet based search techniques should in principle be able to find Higgs signals in our scenarios, provided that the total SUSY production cross section and the fraction of events that contain a Higgs boson are large enough. 

To see this in more detail, we simulate the signal for the points 
\begin{align}
	&({\rm I}) &&M_A = 1000~{\rm GeV} && M_1 = 220~{\rm GeV} && \mu = 280~{\rm GeV} && \tan\beta = 10\,, \notag \\
	&({\rm II}) &&M_A = 300~{\rm GeV} && M_1 = 280~{\rm GeV} && \mu = 400~{\rm GeV} && \tan\beta = 50\,, \notag \\
	&({\rm III}) &&M_A = 300~{\rm GeV} && M_1 = 135~{\rm GeV} && \mu = 400~{\rm GeV} && \tan\beta = 10\,. \notag 
\end{align}
The three points represent the several broad regimes in which one can get rather large Higgs production branching ratios, compatibly with a correct relic abundance\footnote{After the first version of this paper was completed, new results from Xenon~100~\cite{Aprile:2011hi} appeared that exclude point (II) at 90\% CL. This point is however still consistent with the dark matter relic density and the current Xenon~100 constraints for $\tan\beta \sim 30$, for which the collider signatures remain approximately the same.}. Point (I) is representative for the large $M_A$ regime, where sizable Higgs production is obtained for $M_1$ between $150$~GeV and $400$~GeV. The chosen value of $M_1=220$~GeV is not particularly optimized to maximize the production of Higgs bosons, but a good compromise, since larger gaugino masses decrease both the gluino production cross section and the average boost of the Higgs boson. 

Points (II) and (III) are instead representative for the intermediate $M_A$ regime (see Fig.~\ref{fig:combined3}). The first point is away from the resonant region, the second instead lies close to the resonance. As a consequence, in this latter case, we had to choose a rather tuned value for $M_1$ (135 GeV) to obtain a correct dark matter relic abundance. Assuming gaugino universality, this implies a rather light gluino with a mass of around $800$~GeV, which is only slightly above the most recent LHC constraints~\cite{Khachatryan:2011tk,Collaboration:2011qk}.

In addition, we also simulate one point corresponding to a scenario with nonuniversal gaugino masses. From Fig.~\ref{fig:smallm1} we find that the point
\begin{align}
	&({\rm IV}) && M_A = 300~{\rm GeV} && M_1 = 49~{\rm GeV} && M_2 = 400~{\rm GeV} && \mu=300~{\rm GeV} && \tan\beta = 10 \notag
\end{align}
satisfies the relic density constraint, while offering a large Higgs production rate from left-handed squark and stop decays. $M_3$ is fixed to $M_3 = m_{\tilde q} = 1$~TeV. 

\subsection{Higgs Signal Rates at the 14~TeV LHC}

Production of supersymmetric particles at the LHC with $\sqrt{s}=14$~TeV is simulated using Pythia 8, version 8.145~\cite{Sjostrand:2007gs}. The leading order cross sections for squark and gluino production were in addition checked using Prospino~\cite{Beenakker:1996ed}.  Sparticle decays are simulated using decay tables generated with SUSY-HIT~\cite{Djouadi:2007kx}. Higgs decays are switched off to simplify the analysis. 

We impose a very elementary set of cuts, namely we require
\begin{itemize}
	\item ${E\!\!\!/}_T > 200$~GeV,
	\item at least two jets, with $p_{T1} > 300$~GeV and $p_{T2}>200$~GeV. 
\end{itemize}
The missing energy cut serves to suppress SM backgrounds from $Z$+jets and $W$+jets production\footnote{In addition a veto on hard isolated leptons could be used to suppress this background. } and from jet energy mis-measurements in hard QCD events. 
Demanding hard jets also reduces the supersymmetric backgrounds from direct neutralino and chargino pair production which is sizable for small values of $M_1$ and $\mu$.

\begin{table}[t]
\begin{center}
\begin{tabular}{|c|c|c|c|c|}\hline 
 & $\sigma$[pb] &$\sigma_{\rm cut}$[pb] & $\sigma_h$[fb] & $\sigma_\text{boosted}$[fb] \\\hline 
 (I) & 1.11 & 0.52 & 78 & 31 \\\hline
 (II) & 0.73 & 0.34 & 116 &  31\\ \hline 
 (III) & 2.59 & 0.90 & 360 & 135 \\ \hline
 (IV) & 1.60 & 0.83 & 231 & 101 \\\hline
\end{tabular}
\caption{Cross sections for SUSY production at the LHC with $\sqrt{s}=14$~TeV. Shown are the total production cross sections, and the cross sections of events that pass our simple cuts. The fourth column shows the cross section for events that contain at least one Higgs boson, while for the last column we require in addition that the Higgs has a transverse momentum $p_T>200$~GeV.  }
\label{tab:cx}
\end{center}
\end{table}

In Tab.~\ref{tab:cx} we show the total production cross sections and the cross sections for events that pass the basic cuts for the four benchmark points. The fourth column gives the cross sections for events in the cut sample that contains at least one Higgs boson, analogously the last column the cross section for boosted Higgs.

The large gluino mass inhibits larger production cross sections for the first two points. The cross section for events containing a Higgs boson for scenarios (I) and (II) is in fact of order $0.1$~pb, corresponding to 1000 events with 10~fb$^{-1}$. This signal will be challenging to find at the LHC using conventional cut based analyses, but might be possible if one properly makes use of the heavy spectrum of produced particles~\cite{Huitu:2008sa}.

For the jet substructure based analyses to be applicable, at least a fraction of the events must have Higgs bosons with $p_T > 200$~GeV~\cite{Kribs:2010hp}. The transverse momentum distributions of the Higgs bosons in our samples are shown in Fig.~\ref{fig:ptdist}. Points (I) and (II) have ${\cal O}(30~\rm{fb})$ cross sections for boosted Higgs bosons. 

The third scenario (point (III)) has a larger production cross section for sparticles thanks to a smaller gluino mass. Together with a large branching fraction for boosted Higgs bosons, this leads to an enhanced cross section for boosted Higgs bosons.  

The nonuniversal point (IV) has a production cross section after cuts similar to (III), in spite of having a slightly heavier gluino. The reason is that, due to the very light $\tilde{N}_1$, more jets from squark decays pass the cuts. Also note that the fraction of Higgs events with a boosted Higgs boson is larger than for the other benchmark points. This feature is again largely due to the small $\tilde{N}_1$ mass. 

\begin{figure}[t]
\center
\includegraphics[width=.7\textwidth]{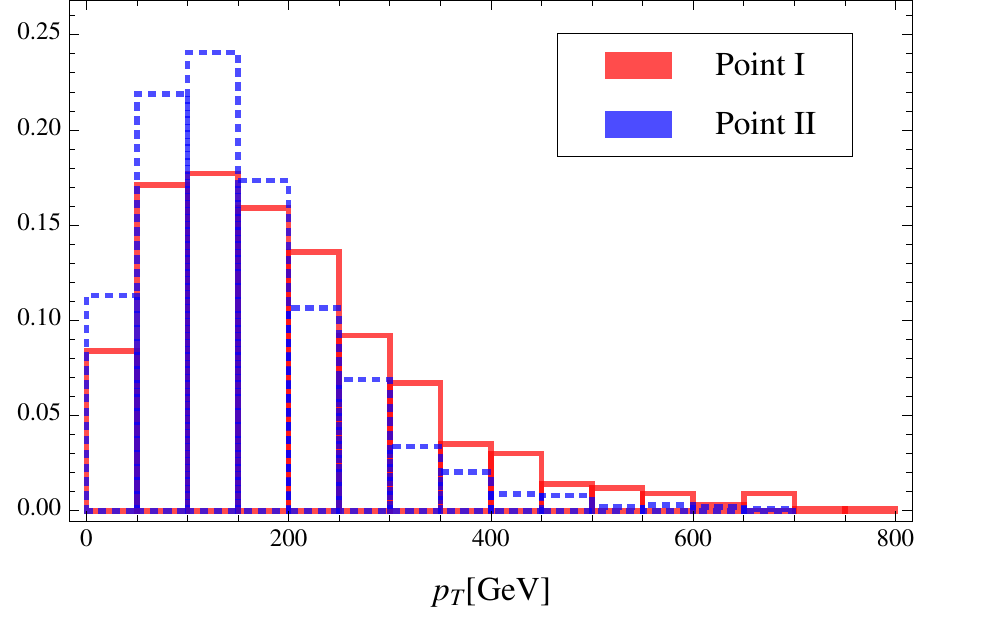}\\
\caption{\small Normalized transverse momentum distributions of Higgs bosons in event samples corresponding to scenarios (I) and (II). The fraction of events with $p_T > 200$~GeV is 40\% and 25\% respectively.
}
\label{fig:ptdist}
\end{figure}
The total cross sections for SUSY cascades with Higgs bosons are comparable to those obtained for the parameter points that were studied in~\cite{Kribs:2010hp}. The fraction of events with strongly boosted Higgs bosons tends to be slightly smaller, around 30-40\% compared to 50\% in~\cite{Kribs:2010hp}, since in our case some of the Higgs bosons originate from longer decay chains.  

Despite the slightly reduced number of boosted Higgs bosons, the similarity  
with the results of~\cite{Kribs:2010hp} suggests that 
 the Higgs boson can be discovered in SUSY decay chains also in the regions where the neutralino relic density agrees with the observed dark matter abundance, with moderate luminosity.
\subsection{Prospects for Higgs Searches at an Early 7~TeV LHC Run}\label{sec:7tev}

Before ending our analysis, we would like to discuss the production of Higgs bosons in MSSM decay chains in the current 7~TeV run of the LHC, and comment on the prospects for observing these events. 
In Tab.~\ref{tab:cx7_1} we show the production cross section at 7 TeV for the same parameter points analyzed in Sec.~\ref{sec:lhc}. Clearly, the large squark and gluino masses inhibit large event rates, such that we can at most expect ${\cal O}(10)$ Higgs bosons per fb$^{-1}$. 

\begin{table}[t]
\begin{center}
\begin{tabular}{|c|c|c|c|c|}\hline 
 & $\sigma$[pb] &$\sigma_{\rm cut}$[pb] & $\sigma_h$[fb] & $\sigma_{\rm boosted}$[fb] \\\hline 
 (I) & 0.092 & 0.019 & 2.7 & 1.1 \\\hline
 (II) & 0.042 & 0.015 & 5.1 & 1.1\\ \hline 
 (III) & 0.113 & 0.030 & 10 & 3.6 \\ \hline
 (IV) & 0.106 & 0.029 & 8.2 & 3.3 \\ \hline
\end{tabular}
\caption{Cross sections for sparticle production at the LHC with $\sqrt{s}=7$~TeV, for squark masses of $1$~TeV. All other parameters are chosen as in Tab.~\ref{tab:cx}.  Shown are the total production cross sections, and the cross sections of events that pass our simple cuts. The fourth column shows the cross section for events that contain at least one Higgs boson, while for the last column we require in addition that the Higgs has a transverse momentum $p_T>200$~GeV. }
\label{tab:cx7_1}
\end{center}
\end{table}

On the other hand, in spite of the recent constraints on the MSSM parameter space coming from LHC, squark masses as low as 800~GeV are still allowed for most of the parameter space~\cite{Khachatryan:2011tk,Collaboration:2011qk}, and it is still relatively easy to find regions where the squark mass can be lowered further, e.g. by reducing the mass gap between the squarks and their immediate decay products. In Tab.~\ref{tab:cx7_2} we show the cross sections for sparticle production for squark masses of $800$~GeV. For point (IV) we have also lowered the gluino mass parameter $M_3$ to $800$~GeV. Since now the hard jets coming from the initial squark decays will have smaller transverse momenta, the cuts on the jet transverse momenta have also been lowered to $200$~GeV and $150$~GeV for the hardest and second hardest jet, respectively. 

\begin{table}[t]
\begin{center}
\begin{tabular}{|c|c|c|c|c|}\hline 
 & $\sigma$[pb] &$\sigma_{\rm cut}$[pb] & $\sigma_h$[fb] & $\sigma_{\rm boosted}$[fb] \\\hline 
 (I) & 0.23 & 0.086 & 11 & 3.0 \\\hline
 (II) & 0.18 & 0.063 & 17 &  2.0\\ \hline 
 (III) & 0.31 & 0.142 & 36 & 11 \\ \hline
 (IV) & 0.36 & 0.169 &  45 & 14 \\ \hline
\end{tabular}
\caption{Same as Tab.~\ref{tab:cx7_1}, for squark masses of $800$~GeV, and with jet $p_T$ requirements relaxed to $200$~GeV and $150$~GeV respectively. 
For point (IV) also the gluino mass parameter $M_3$ has been lowered to $800$~GeV.}
\label{tab:cx7_2}
\end{center}
\end{table}

The cross sections go up by roughly a factor of three, if compared to the case $m_{\tilde Q}=1$ TeV. An additional NLO K-factor of $\sim 1.3$ should be applied to these results~\cite{Beenakker:1996ed}. 

The downside of lowering the squark masses is that a smaller fraction of Higgs bosons satisfies the boosted criterion. In particular the benchmark points (I) and (II) suffer from this effect, when comparing with the case of $1$~TeV squarks. Points (III) and (IV) are less sensitive, since, for these points, a large part of the Higgs boost comes from the mass difference between the lightest and the heavier neutralinos that is not affected by the reduced squark masses. 
For point (III) and (IV) we expect respectively roughly 11 and 14 boosted Higgs events per experiment at the end of 2011, which might be sufficient to observe an excess in the boosted discovery channel.

While a more detailed analysis is required to determine whether these events can be observed at this early stage, the event rates (at least) for points (III) and (IV) give rise to some hopes. Clearly these points are also the most constrained scenario and at the point of being probed by the LHC experiments. The most recent constraints from ATLAS~\cite{Collaboration:2011qk} actually exclude squark masses $m_{\tilde q} \lesssim 800$~GeV for gluino masses $m_{\tilde g} \approx 800$~GeV, however this analysis assumes a very simplified spectrum with a massless LSP. 
The parameter point (III) is better approximated by MSUGRA with $M_{1/2}=335$~GeV and $m_0=375$~GeV, which is still allowed~\cite{Collaboration:2011qk}, and reproduces the physical squark and gluino masses of point (III). 
\section{Conclusions}\label{sec:conclusions}

In this article, we have analyzed the possibility of observing Higgs bosons proceeding from the decay of supersymmetric particles in regions
of parameter space consistent with the observed neutralino relic density.  For this purpose, we have not analyzed any particular 
realization of the MSSM, but we have concentrated on the low energy properties of the model, studying its dependence on the gaugino and
higgsino masses, as well as on the non-standard Higgs, squark and slepton parameters at the EW scale. 

Relatively light squarks and gluinos tend to increase the cross section, but they lead to a suppression of the fraction of boosted Higgs bosons;
the prospects for observation in boosted Higgs searches are therefore not very sensitive to the exact mass scale,  provided it is about a TeV.  Taking the standard
mass unification relation for the gaugino masses,  in the region consistent with the observed dark matter density, Higgs bosons proceed
 mainly from the decay of the heaviest chargino, second lightest and heaviest neutralino. A considerable fraction of all squark and gluino decays contain 
 Higgs bosons and a sizable fraction of them are highly boosted, implying  a good prospect for observation even, perhaps, 
 at the 7~TeV machine after combination of the ATLAS and CMS data. 
 
 Prospects for Higgs observation become weaker if there are light sleptons in the spectrum. The reason is that, in such a case, the
 chargino and second neutralino decays may be dominated by decays into slepton-lepton pairs, diminishing the possibility
 of Higgs observation. 
 
 We have also analyzed the variation of Higgs production for non-universal gaugino mass parameters. In general, the production of the SM-like Higgs boson may not be enhanced much in 
 such conditions, but production of non-standard Higgs bosons may be highly enhanced, reaching up to 25\% for certain gaugino spectra. An additional virtue of non-universal gaugino masses is that more parameter space becomes available at small $M_1$, and we have found that also there a large fraction of supersymmetric events contain light Higgs bosons.
 
 To finalize, let us stress the relevance of searching for Higgs bosons in cascade decays of heavy supersymmetric particles. At a minimum, it will provide the possibility of detecting a SM-like Higgs boson decaying into bottom quarks, its dominant decay channel.  This production channel will  be complementary to the standard search for boosted Higgs bosons; the relative strength of these search channels will depend on the supersymmetric spectrum.    In addition, for low values of $M_A$, for which the significance of the standard gluon fusion and weak boson fusion Higgs boson search channels is weakened (see for instance, Ref.~\cite{Draper:2009au}), it could serve as a discovery channel in an early LHC run.  For larger values of $M_A$, depending on the values of the squark and gluino masses, it could still serve as a competitive search channel (for a more detailed discussion, see e.g. Ref.~\cite{Kribs:2010hp}). Detecting a SM-like Higgs boson in several channels will be of central importance in order to determine its production and decay properties, and therefore to understand the mechanism of electroweak symmetry breaking that leads to the generation of mass of all known elementary particles.  

~\\
{\bf Acknowledgements} \\

The authors would like to thank G. Kribs and A. Martin for useful discussions. 
 Work at ANL is supported in part by the U.S. Department of Energy (DOE), Div.~of HEP, Contract DE-AC02-06CH11357. 
 P.S.  is also partially  supported  by the UIC DOE HEP Contract DE-FG02-84ER40173. We would like to thanks the Aspen Center for Physics, where part of this work was performed.
\addcontentsline{toc}{chapter}{Bibliography}
\bibliographystyle{utphys}
\bibliography{bib}
\end{document}